\documentclass[sigconf]{acmart}

\usepackage{booktabs} 
\usepackage{subfig}
\usepackage{url}
\usepackage[inline]{enumitem}
\usepackage{multicol}

\setcopyright{none}

\acmDOI{}

\acmISBN{}

\acmYear{}

\acmArticle{}
\acmPrice{}

\acmBooktitle{}
\editor{}
\editor{}
\editor{}

\begin{document}
\title{A Spin-based model checking for the simple concurrent program on a preemptive RTOS}

\author{Chen-Kai Lin}
\affiliation{%
  \institution{Institute of Information Science, Academia Sinica}
  \city{Taipei}
  \country{Taiwan}
}
\email{kai.zsv@gmail.com}

\author{Ching-Chun (Jim) Huang}
\affiliation{%
  \institution{Department of Computer Science and Information Engineering, NCKU}
  \city{Tainan}
  \country{Taiwan}
}
\email{jserv@ccns.ncku.edu.tw}

\author{Bow-Yaw Wang}
\affiliation{%
  \institution{Institute of Information Science, Academia Sinica}
  \city{Taipei}
  \country{Taiwan}
}
\email{bywang@iis.sinica.edu.tw}
\renewcommand{\shortauthors}{}

\begin{abstract}
We adapt an existing preemptive scheduling model of RTOS kernel by eChronos from machine-assisted proof to Spin-based model checker. The model we constructed can be automatically verified rather than formulating proofs by hand. Moreover, we look into the designs of a Linux-like real-time kernel--Piko/RT and the specification of ARMv7-M architecture to reconstruct the model, and use LTL to specify a simple concurrent programs--consumer/producer problem during the development stage of the kernel. We show that under the preemptive scheduling and the mechanism of ARMv7-M, the program will not suffer from race condition, starvation, and deadlock.
\end{abstract}


\maketitle

\section{Introduction}
Model checking \cite{Clark:2000, Clarke:2008:BMC:1423535.1423536} is first introduced by Edmund Clarke and Allen Emerson in 1981 to solve the Concurrent Program Verification. It has a state-transition graph (denote as a model) and some logic formulas (denote as properties). The algorithm of model checking will try to claim that all paths of the model are satisfied the properties. In other words, if we write a property about mutual exclusion and the model checker says the property is unsatisfied, we can trace the error messages back to the situation and inspect whether the model or the real-world system will suffer from the race condition.

Concurrent errors are hard to detect by code review or test case, because we can not guarantee the asynchronous processes executing in a specific order. Another way to ensure the correctness of concurrent programs is by proof, one example is that Lamport proves the correctness of a fast mutual exclusion algorithm in 1987 \cite{Lamport:1987:FME:7351.7352}. But, this strong guarantee requires the preliminary of proof tactic and constructs by hands. Model checking, though, has weaker guarantee but is more expeditious only if the abstract model and the real-world system are homogeneous. Furthermore, the correctness of the algorithm can not stand for the correctness of the implementation on other hardware or software mechanisms, i.e. interrupt or softirq. Holzmann \cite{Holzmann:2014:MC:2556647.2560218} indicates model checking is one of the three topics that the Mars Science Laboratory (MSL, launched by NASA) has used to reduce risk in complex software systems. He also points out that they had successfully verified several concurrent issues on key parts of the spacecraft, including Cassini, Deep Space One, and the Mars Exploration rovers.

Piko/RT \cite{PikoRT}, developed at National Cheng Kung University, is a non-trivial operating system but small enough for verification purpose. One of the key features is real-time capability. It enables interrupts (exceptions) almost all the executing time to handle the incoming IRQs. Moreover, it is optimized for the ARM Cortex-M series. Until 2015 \cite{ARMProduct}, ARM had shipped 20.6 billion Cortex-M units within six years and the number is growing. It is hard to debug if such a number of devices deployed around in different areas. Not to mention the software system must not behave unexpectedly in the critical usages.

In this paper, we address the problem of verifying the simple concurrent programs will not act unexpectedly (no race condition, starvation, and deadlock) under the specific conditions (preemptive scheduling and interrupt-driven mechanisms). We first construct an abstract model on the Spin model checker based on three domains: proof framework of eChronos, source code of Piko/RT, and reference manual of ARMv7-M, discussing in section 3. Next, write three properties (race condition free, starvation free, and deadlock free) and verify in Spin (section 4 and 5). Last but not least, discuss the background knowledge and summarize the conclusions in section 2 and 6. All our model\footnote{\url{https://github.com/kaizsv/pikoRT-Spin.git}} is in the public domain.

\section{Background}
We use Spin as our model checking tool and build an abstract scheduling model on. In this section, we discuss some topics about Spin model checker and the real-time scheduling before approaching to the model. And then, talk about several related works of this paper.

\subsection{The model checker}
\subsubsection{Spin}

\begin{figure}
\subfloat[][Finite state infinite run automata.]{\includegraphics[width=0.22\textwidth]{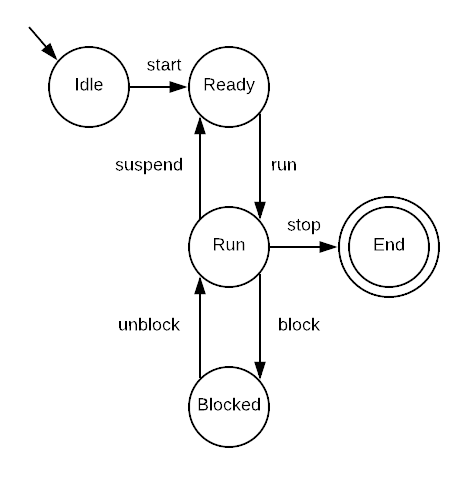}\label{fig:fsaex}}
\quad
\subfloat[][A path of $\omega$-run]{\includegraphics[width=0.16\textwidth]{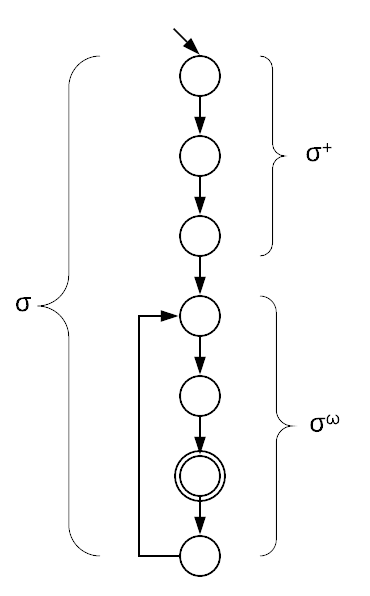}\label{fig:buchi}}
\caption{Example of $\omega$-automaton}
\end{figure}
Spin \cite{SpinRoot, Ben-Ari:2008:PSM:1349767, Holzmann:2003:SMC:1405716} is an open-source verification tool of multi-threaded software developed in the Computing Science Research group of Bell Lab. The theory behind Spin is the $\omega$-automaton, which is the variation of classic finite state automata with acceptance conditions on infinite executions. Consider Figure \ref{fig:fsaex}, note that the termination of the automata would not necessarily be a desirable result. We can declare some state that will be visited infinitely often by labeling \textit{accept}, i.e. Run state, as a legal run. Because the states are finite, the infinite run ($\omega$-run) will anyway repeat itself at the certain degree just like Figure \ref{fig:buchi} shows. If Spin find an $\omega$-run with some state labeling \textit{accept}, it can claim that the automata is accepted. This is so called the B\"uchi acceptance.

Every thread in Spin is an automata and product altogether in the verifying stage. Spin supports a meta language: PROMELA, which interpreting readable script into automata, like common condition statements, \texttt{if} or \texttt{do}. It also provides \texttt{channel} allowing the communication between two threads. And the most important one in this paper is the \textit{guard}. \textit{guard} is a true of false expression that guard the execution of the following statements only if it is true, otherwise the thread will be blocked until it becomes true.

\subsubsection{Linear temporal logic}
\textit{Safety and liveness}. We need to discuss these two kinds of properties before going to linear temporal logic, which are introduced by Lamport in 1977 \cite{Lamport:1977:PCM:1313313.1313439}. Safety property guarantees nothing bad ever happens, e.g., more than two processes access same critical section at the same time will never happen. Liveness property denotes as something good eventually happens, e.g., if a process wants to enter its critical section, it can eventually enter. Above is only the informal expression of two types of properties.

\textit{Fairness} is another issue of temporal logic. We can not assume relative speeds of asynchronous processes, there might exist a run that when a process can execute, but never be executed. This is called the unfair processes. Two commonly used variants of fairness are:

{\medbreak\indent\textit{\textbf{weak fairness}: If a statement is continuously enabled, it will eventually be executed}}
{\\\indent\textit{\textbf{strong fairness}: If a statement is enabled infinitely often, it will eventually be executed}}
\medbreak

\begin{figure}
\includegraphics[width=0.9\linewidth]{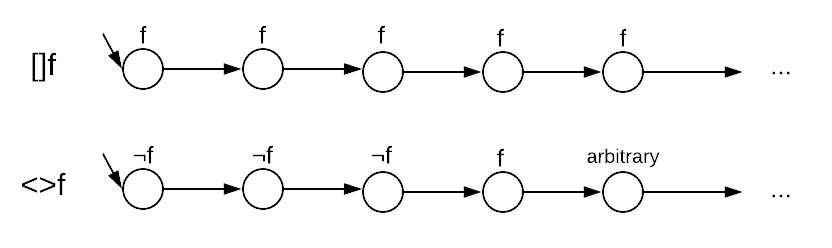}
\caption{Semantics of Global and Future operators}
\label{fig:ltl}
\end{figure}

Temporal logic allows us to formalize the properties of a run unambiguously with some special operators. Most relevant to the verification of asynchronous process systems is a specific branch of temporal logic, \textit{linear temporal logic} (LTL). LTL is sufficient to describe two operators in the scope of this paper. Let $f$ be a LTL formula. Operators in $f$ such as \texttt{not}, \texttt{and}, \texttt{or}, \texttt{implication}, and \texttt{equivalence} are still a LTL formula. Two further temporal operators \textit{global} and \textit{future} are introduced below and Figure \ref{fig:ltl}:

{\medbreak\indent\textit{$\Box f$: $f$ is true now and forever in rest of the run}}
{\\\indent\textit{$\Diamond f$: $f$ is true eventually in the future run}}
\medbreak

\begin{figure}
\includegraphics[width=0.7\linewidth]{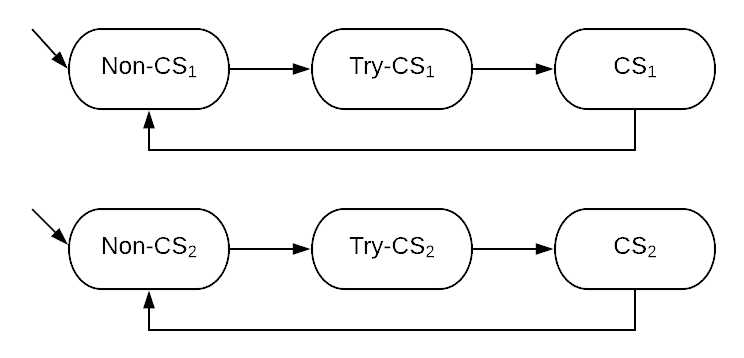}
\caption{Mutual exclusion}
\label{fig:mutex}
\end{figure}

Let us consider an example of mutual exclusion. Figure \ref{fig:mutex} contains two processes and \textit{CS} is the critical section of that process. Each process is first in non-critical section. If one process wants to access its shared variables in critical section, it must try to gain lock first at \textit{Try-CS} state and leave the critical section as soon as possible. We can write down the LTL formula about mutual exclusion free, starvation free, and deadlock free as follows:

{\medbreak\indent\textit{\textbf{mutual exclusion free}: $\Box\neg(\text{CS}_1\wedge\text{CS}_2)$}}
{\\\indent\textit{\textbf{starvation free}: $\Box\Diamond \text{Try-CS}_1\rightarrow\Box\Diamond\text{CS}_1$}}
{\\\indent\textit{\textbf{deadlock free}: $\Box\Diamond(\text{Try-CS}_1\wedge\text{Try-CS}_2)\rightarrow\Box\Diamond(\text{CS}_1\vee\text{CS}_2)$}}
\medbreak

The first property is simple. Both processes can not access its critical section at the same time in every state of a run. The rest are the liveness properties. Starvation free means if a process try to enter its critical section infinitely often, it will be in critical section infinitely often. And deadlock free property means if both processes try to enter each critical section infinitely often, at least one process can enter its critical section infinitely often. Note that ``infinitely often'' represents the properties are strong fairness. This is will be discuss later in section 4.

To understand how Spin verifies LTL formula, consider it as a game. Spin negates the LTL formula and converts it into \textit{never claim}. The \textit{never claim} is also an automaton that can product together with the model. The Spin model checker then try to find a path that satisfied the negated formula which means there exists a counterexample in the model that violates the property. On the contrary, if \textit{never claim} never reaches the acceptance state (never finds a path that violates the formula) means the model satisfied the formula.

\subsection{Real-time scheduling}
A real-time system is a set of tasks which running interleaving on uniprocessor or concurrently on multi-processors that can simply classify into preemptive and non-preemptive scheduling. Non-preemptive scheduling has lower context switch latency but the interrupt handler might violate the deadline. And preemptive scheduling forces each task to process under a limited time and the task might be blocked (pending) even if within its timeslice. In this section we discuss several factors which affect the behavior of preemptive scheduling from the hardware and software aspects. The hardware is the mechanisms of ARMv7-M architecture and the software is the implementation of the case study.

\subsubsection{Exception of ARMv7-M architecture}
ARMv7-M architecture \cite{ARMv7-M:Manual, Yiu:2013:DGA:2602039} uses ``exception'' as a response to system event and ``interrupt'' as a peripheral request. Exception (or interrupt) with higher priority level can preempt the lower one. Moreover, synchronous exception like non-maskable interrupt (reset, hardware fault), system fault, and SVCall must execute before the context switch. This means synchronous exception can only be preempted by higher priority level exception before changing to the next task. The incoming exception will be pending if the priority level is less than or equal to the current one. Other exceptions are considered asynchronous that we can not guarantee the timing of executing, even after the context switching.

\textit{Exception entry and exception return} are the mechanism before and after the exception handler take place on ARMv7-M architecture. While an exception occurs, the hardware will compare the priority of current task and the coming exception. The exception entry will push parts of the context into the stack and reset return values of current context if the coming exception is allowed to preempt according to the interrupt policy. After finishing the handler task, the exception return will choose a proper task (from the stack or the pending state) to switch to.

\textit{Tail-chaining} is the further extension of the exception return. Tail-chaining can continue to process the pending exception at the exception return stage. Rather than restoring to the return context from the stack and saving it back again to the stack, tail-chaining can directly switch to the pending exception with less timing gap.

\textit{Context switch} is a significant issue in this paper. Context switch typically requires processor to execute at a critical region of interrupt disabled to avoid the corruption of data during the switch, However, the busy waiting might cause pending the coming exception and violate the deadline. ARMv7-M supports interrupt enabled context switch. SVCall is a synchronous exception and triggered by the \texttt{svc} instruction from user tasks to perform supervisor call. PendSV is an asynchronous software interrupt and handles the scheduling point enabled by the systick handler. To enable interrupt during context switch, the ARMv7-M reference manual suggests to configure both exceptions with the lowest priority level so that they can not preempt each other.

\subsubsection{Preemptive scheduling}
Preemptive scheduling causes the task to stop and resume executing frequently. One reason is the task reaches its timeslice and does not finish yet, the scheduler need switch to the next task from the runqueue. Another reason is when a task with higher priority added into the runqueue, at any scheduling point the system needs to guarantee the priority of current task is always the highest one in the runqueue. The following lists several aspects which will influence the behavior of the preemptive scheduling.

\textit{Bitmap scheduler} has 32 priority levels and two priority arrays, active and expired, which can be swapped in constant time. The scheduler always chooses the highest priority task from the active priority runqueue and selected in round-robin way if multiple tasks exist in the same priority level. The implementation of array data structure and the bitwise operations make time complexity in constant time.

\textit{Softirq and tasklet}. Interrupt handlers are asynchronous and often timing-critical so there are several limitations, such as respond rapidly and cannot be blocked. Softirq is the mechanism that moves the execution of the non-critical job (bottom half task) from interrupt handlers to the user task (softirq process). Moreover, softirq maintains a bitmap scheduler with only one runqueue to manage the different bottom half tasks, called tasklet.

\textit{Mutual exclusion} is one of the solutions to prevent race condition and has different implementations. In this paper, we focus on a pair of load-link/store-conditional instructions, \texttt{ldrex} and \texttt{strex}, supported by a state machine \textit{exclusive monitor} provided by ARMv7-M architecture. Each processor has a exclusive monitor called local monitor (we only consider uniprocessor situation here), which will mark a segment of memory address (register) as \textit{exclusive} by the calling of \texttt{ldrex} instruction. One local monitor can only mark one register at the same time, \texttt{ldrex} will load a value from the address to the register and clear the previous mark and mark the address as exclusive. \texttt{strex} conditionally stores the value back to the address if the address is marked as exclusive and clear the mark at the same time, otherwise the store will fail notifying by the return value. In brief, no matter how many times \texttt{ldrex} mark the address, only one can update the value of that address by \texttt{strex}. And finally, all mark will be cleared at every exception return stage. This make sure the atomic operations of the mutex variables.

\textit{Condition variable} is an extension of mutual exclusion. It will temporary release the lock and move itself out of the runqueue if the condition is not satisfied (condition wait). While others get the lock and satisfy the condition statement, it signals the blocked one by returning the lock and enqueuing the blocked one again (condition signal). This is more efficient than busy waiting because the blocked one will temporarily gives up the processor's resources and turns to the process.

\subsection{Related work}
\textit{eChronos} \cite{DBLP:conf/itp/AndronickLMMR16} is a tiny real-time OS running on ARM Cortex-M3 platform developed by Data61 in Australia. They provide a framework and prove the property ``the running task is always in the highest priority'' on the Isabelle/HOL proof assistant based on the Owicki-Gries \cite{Owicki:1976:APT:2696887.2697004} method. The OG method is the parallel version of Hoare logic, known as the pre- and post- condition style, guaranteeing the shared variables are not interleaving by the parallel programs. Their framework is delicate and scalable that simulates both software and hardware behaviors in one model. But the challenge of their work is the exhausting hand constructed proof and the lack of implementation details. The reason of the later one is the proof strategy that they do not declare the explicit interrupt policy, causing all exceptions can preempt each other. Instead, they have proved a larger range of the exception behavior is correct that can implicate to the limited and real one.

\textit{OSEK-Spin}, Zhang, et al. \cite{10.1007/978-3-319-17581-2_16} use Spin to verify the application of a standard automobile OS, OSEK/VDX, which is widely adopted by many automobile manufacturers. To enhance the user experience, different applications are developed based on the OS. To check whether the applications can run correctly, the researcher provide a Spin-based model that can replicate the executions on OSEK. The correctness means under the concurrent running task, synchronous event, and context switch, the user application will not suffer from race condition or priority inversion. They also developed bounded model checking to verify more complex applications. However, the model considered only the software aspect and did not contain the specific hardware behavior.

The two researches list above are the most related to our paper. Moreover, there are lots of corresponding work about verifying software or hardware by mathematical proof, for example, seL4 \cite{Klein:2009:SFV:1629575.1629596}. However, the design of seL4 kernel is optimized for the formal verification, there are some limitations. Although seL4 had analyzed the WCET \cite{6121451}, its kernel code only has few interrupt points, others with interrupt disabled. This is different from the common design of RTOS.

\section{The model}

\begin{figure*}
\includegraphics[width=0.7\linewidth]{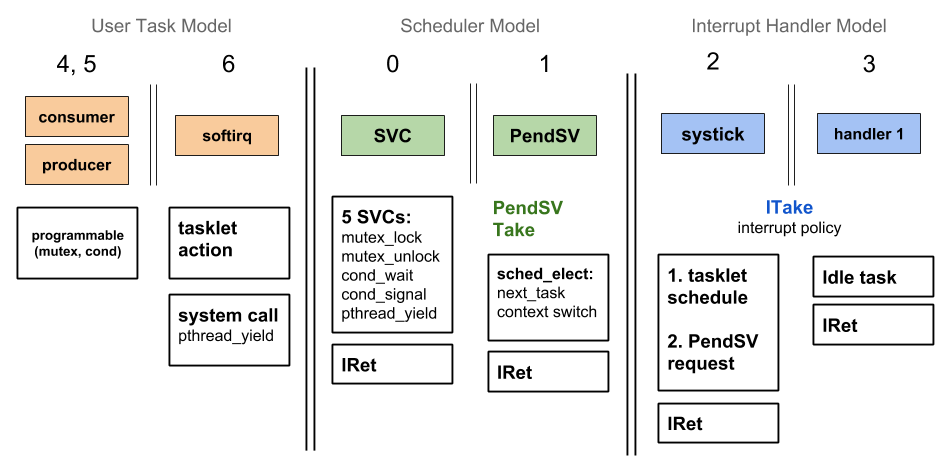}
\caption{The schema of the model (2 user tasks, 2 interrupts)}
\label{fig:model}
\end{figure*}

In section 2, we had discussed several factors of scheduling. In this section, we show how we model those factors by PROMELA. The model consists of three static processes (SVC, PendSV, and softirq) and an amount of user tasks and interrupts (number of interrupt must contain one systick). Figure \ref{fig:model} is a schema of 2 user tasks and 2 interrupts. The model can separate into three parts: user task for the orange box, exception for the blue one, and two special exceptions SVC and PendSV for the green one. Each process has a PID listed at the top of Figure \ref{fig:model}.

\subsection{Overview}
We had mentioned that each process are executed in asynchronous order, we use this feature to simulate the incoming of the IRQs. But to determine which process is the legal one to occupy the resource of processor, eChronos introduced a guard called \textit{AWAITS} to control the executing of the processes. Another variable \textit{AT} (active task) records which process is the legal on to be executed. Every statement in the model will be wrap by \textit{AWAITS} and can be executed only if the \textit{AT} becomes its PID, otherwise, the statement will be blocked until AT becomes its PID again. There is also an \textit{A\_AWAITS} statement which do the same thing as \textit{AWAITS}, because there are two atomic statements in PROMELA, \texttt{d\_step} and \texttt{atomic}. Both statements guarantee no other processes can interleave within the scope of the statement. The difference is that in verification stage, the \texttt{d\_step} is seen as one step while the \texttt{atomic} is depend on the statements inside. Note that each statement in the code segment surrounding by a frame is wrapped by the \textit{AWAITS} or \textit{A\_AWAITS} statements which are the unit operation in our model.

Starting from SVC handler first, we modeled five system calls, including, mutex lock, mutex unlock, condition wait, condition signal, and pthread yield. Consider that SVC is synchronous, we use a synchronous channel called rendezvous channel in PROMELA to perform the behavior of calling the system call. It is something like the client/server model that the SVC handler is the server and serves several user tasks. The difference is that other processes remain blocked during the communication. Recall the interrupt enabled context switch, PendSV is asynchronous and used to respond the scheduling request from the systick. Simultaneously, the systick handler will insert its bottom half task into the tasklet runqueue waiting to be executed in softirq process. Another interrupt handler do nothing in the model. Last, we choose the classic consumer/producer program as our application to verify and the softirq will discuss later.

\subsection{Exceptions of ARMv7-M architecture}
Figure \ref{fig:model} shows that every exception is starts with \textit{ITake} or \textit{PendSVTake} and ends with \textit{IRet} (except SVC is synchronous) which models exception entry and exception return. There are two situations in \textit{ITake}. First, we refer to the manual of ARMv7-M to implement the interrupt policy, including the priority and the pending state to determine whether the incoming IRQs can preempt the current process. \textit{PendSVTake} is a simplified \textit{ITake} and only affects to the PendSV process. An array \textit{ATStack} stores the preempted processes that has not finished its job yet. \textit{IRet} elects the next running task from the top element of \textit{ATStack} or the highest priority of pending state when the current interrupt handler is finished. If the exception priority of the top of the \textit{ATStack} is higher, pop the exception directly to the \textit{AT}. If not, remember that the exception in pending state has not been executed yet, an additional variable \textit{ghost\_direct\_AT} is introduced to record this situation and can be detected by the second condition of \textit{ITake}. The choosing of pending state running first is similar to the tail-chaining mechanism in ARMv7-M

The design of \textit{ATStack} is an array data structure that the last element in it must be user task, others are exceptions in priority order. The formalization of context switch is to change the last element in \textit{ATStack}. Because context switch only occurs in SVC or PendSV processes, the only element in \textit{ATStack} must be current user task and the \textit{AT} must be the PID of SVC or PendSV. We add several assert to check this scene and change the last element in \textit{ATStack} to reproduce the behavior of context switch.

\subsection{Preemptive scheduling}
Two priority runqueues, which contain an unsigned map and a queue, form the structure of bitmap scheduler. One key difference between the model and the source code is the swap of two priority runqueues. The source code uses pointer to swap each other, however, the model uses bitwise operations to do that. Two priority runqueues \textit{sched\_bm[2]} are defined in the model. Another global bit \textit{SCHED\_STATE\_SWAP} is helped to distinguish each runqueue. To swap the runqueues, just \texttt{xor} the swap bit. Additionally, the \textit{ACTIVE} and \textit{EXPIRED} thread state are designed in the same way.

{\medbreak\indent\textit{\textbf{ACTIVE}: $(0 | \text{SCHED\_STATE\_SWAP})$}}
{\\\indent\textit{\textbf{EXPIRED}: $(1 \oplus \text{SCHED\_STATE\_SWAP})$}}
\medbreak

In short, the \textit{softirq} is a user process which executes the bottom half task of the interrupts. And the \textit{tasklet} is a one priority runqueue managing which bottom half tasks take place first. If no more bottom half task can be executed, the softirq process will give up the processor's resource by calling the pthread yield system call. But notice that if the softirq has higher priority than other user processes, the system call still choose softirq as next process.

The lock and unlock of \textit{mutex} can split into two stages: non-privilege and privilege because of the supporting of load-link/store-conditional (LL/SC) instructions. Take mutex lock for example, if the lock is free then gain lock directly in the non-privilege mode by the mechanism of LL/SC, otherwise, calling the mutex lock system call to give up the execution (moving itself out of the runqueue). Another point to note that, there must have a place to store the tasks moving out of the runqueue by mutex. The implementation of the source code is a linked list structure, we use an array structure in PROMELA, instead. However, to reduce the size of the model, there is only one slot in the array currently and scalable depending on the applications.

Implementation of \textit{condition variable} is based on the mutex that will temporarily release the lock and gain lock again if the condition is satisfied. Nonetheless, this is difficult to implement in PROMELA. In source code, every process performing context switch in SVC or PendSV handlers which will stop at a specific point in the handler and resuming when switching back. If we want to emulate this behavior in PROMELA model, the SVC and PendSV handler must duplicate times of the number of user task. This is a great effort for model checking that we need to share the SVC and PendSV handler within each user task. This leads to another problem that we can not place two scheduling points in one exception handler and the scheduling point must establish just before the \textit{IRet}. This limits the modeling of condition wait system call that we need to use two system calls (condition wait and mutex lock) and wrap together with \textit{atomic} statement to complete the one system call's job.

\section{Strategy of verification}
\begin{table*}
\caption{Options of Spin model checker}
\label{tab:spinv}
\begin{tabular}{rccc}
\toprule
\multicolumn{4}{c}{Spin Version 6.4.8 - 2 March 2018} \\
\midrule
 & \textbf{Spin options} & \textbf{Verification options} & \textbf{Runtime options} \\
 \cmidrule{2-4}
Safety properties & \texttt{-a} & \texttt{-DXUSAFE}, \texttt{-DCOLLAPSE}, & \texttt{-m100000000} \\
& & \texttt{-DSAFETY}, \texttt{-DNOCLAIM}, & (\texttt{-m<changeable>}) \\
& & (\texttt{-DMA=24}), \texttt{-DNOFAIR}, & \\
& & \texttt{-DMEMLIM=<changeable>} & \\
\cmidrule{2-4}
LTL acceptance & \texttt{-a}, & \texttt{-DXUSAFE}, \texttt{-DCOLLAPSE}, & \texttt{-m100000000}, \\
cycles & \texttt{-DLTL} & (\texttt{-DMA=24}), \texttt{-DNOFAIR}, & \texttt{-a}, \\
& & \texttt{-DMEMLIM=<changeable>} & \texttt{-N <ltl name>} \\
\bottomrule
\end{tabular}
\end{table*}
We had talked about the model, now discussing the properties and how we verify the model by Spin model checker.

\subsection{The properties} These two methods are used to verify the correctness of the model:
\begin{enumerate*}[label={\roman*)},font={\bfseries}]
\item \textit{assert} for safety property,
\item \textit{LTL} for liveness properties.
\end{enumerate*}
The \textit{assert} is not only the safety property but also a useful tool to check whether the model is correct during the process of modeling. We use \textit{assert} in the following ways.
\begin{itemize}
\item The insertion of the queue will not induce buffer overflow and the deletion of the queue must succeed.
\item If the value of map is not zero, there must be some elements in the queue.
\item If the scheduler chooses idle as next thread, the current thread must not be idle.
\item The modification of IRQ pending state must be the interrupt process.
\item The tail-chaining only happens when no other exception priority in pending state is higher than itself.
\item The \textit{ATStack} will not have buffer overflow.
\item PendSV can only preempt the user task.
\item The consumer/producer will not suffer from race condition using two global bits \textit{cs\_c} and \textit{cs\_p} to record the entrance of critical section.
\item Only user tasks can perform the system call when there is no pending exception.
\item Only user tasks can perform context switch when there is no other active exception (excluding SVC and PendSV).
\item The last element in \textit{ATStack} is a user task; others are \textit{UNKNOWN}\footnote{ARMv7-M resets the register with UNKNOWN value} during the context switch.
\item Mutex list (array) must not be empty if the value of mutex is larger than zero.
\item The value of mutex must not smaller than -1.
\end{itemize}
For LTL, we label the condition loop as \textit{want}, it means that the process wants to enter its critical section but blocked at the condition loop. The $@$ symbol means the execution is at the statement with specific label. Note that we apply the strong fairness in the LTL formula. If not, the model checker will consider the situation that the systick preempt itself forever. But this is impossible that the duration of systick is 1ms in the real-world system.

{\smallbreak\indent\textit{\textbf{consumer starvation free}: $\Box\Diamond\text{consumer@want} \rightarrow\Box\Diamond\text{cs\_c}$}}
{\\\indent\textit{\textbf{producer starvation free}: $\Box\Diamond\text{producer@want} \rightarrow\Box\Diamond\text{cs\_p}$}}
{\\\indent\textit{\textbf{deadlock free}: $\Box\Diamond(\text{consumer@want}\wedge\text{producer@want})\rightarrow\Box\Diamond(\text{cs\_c}\vee\text{cs\_p})$}}

\subsection{Spin options}

The hardware environment in this paper is a 1TB memory machine with Xeon 6134M 3.2GHz processors. Table \ref{tab:spinv} is the Spin options for Safety and LTL verifications. The search algorithm in this paper is the DFS. Looking at \textbf{Spin options} first, \texttt{-a} is needed to generate the verification code and \texttt{-DLTL} is a custom macro that used to load the LTL properties. In \textbf{Verification options}, \texttt{-DXUSAFE}, \texttt{-DCOLLAPSE}, and \texttt{-DNOFAIR} are used in both properties, the first one is to disable the usage of xr and xs assertions, the second one is to compress each state to reduce the memory usage, the third one is to disable the weak fairness provided by Spin (we use strong fairness in LTL formula). \texttt{-DNOCLAIM} is to disable the usage of never claim, but enables while verifying the LTL properties. \texttt{-DSAFETY} is used in safety property only. \texttt{-DMEMLIM} is the maxima memory that Spin can use and \texttt{-DMA} is the algorithm of minima automata which saving the memory but spending lots of running time. Next, the \texttt{-m<changeable>} is the maxima depth for the DFS algorithm and \texttt{-N} used to choose which LTL property to verify.

Spin model checker will provide statistics if the verification is completed under the memory and maxima depth boundary. It also provides a list of unreached statements in the model but never be executed. To analyze this list, there are eight scheduling points in the model. We copy the macro of scheduling points eight times with different names. With this setting, we can tell the difference between the eight scheduling points by the different lines of code and analyze them later.

\section{Discussion}

We verify a classic concurrent program--consumer/producer in this paper and use a bit to denote the data buffer. This application is also one of the test case of Piko/RT (cond\_3). Table \ref{tab:verification} shows the results of the four verifications. Note that the \textbf{Safety} case uses only \textit{assert} to verify race condition, which is more efficient than other liveness cases. Because the previous one only checks the specific variable at the point, however, the \textit{LTL} checks the variables along every $\omega$-runs (Figure \ref{fig:buchi}). Moreover, the meaning of \textbf{depth} and \textbf{states} in Table \ref{tab:verification} are that the former represents the longest depth of the $\omega$-run and the later is the number of states that Spin had verified. The state is the set of all global variables in the model, each statement in the model might causes the migration from one state to another.

The most interesting point is that the usage of memory is extreme high, is it really worth it to consume such effort to build the verifications? Remember we had mentioned about \textit{AWAITS} statement, it is the watershed separating the atomic and concurrent executing. Between each \textit{AWAITS} (or \textit{A\_AWAITS}) statement is an interrupt point that allowed every non-blocked process to be executed in different order. There are total 67 \textit{AWAITS} (and \textit{A\_AWAITS}) statements in our model, and this is not counting the repeated ones (the \texttt{inline} function of PROMELA will be expended like macro in C language). For example, there are nine scheduling points in the model (8 in SVC and 1 in PendSV) which means we have to multiple nine times the number of interrupt points under the \texttt{inline} function of scheduling point. It is hard for us to consider such amount of scheduling points at the same time by code review or test cases, how can we check a vulnerability at the depth of 50 million steps? However, with the help of the Model Checking and the powerful machine, we can verify those properties within sustainable time. All we have to do is to construct an elegant model which can use less steps to represent the real-world system.

\begin{table}
\caption{The 4 verification results}
\label{tab:verification}
\begin{tabular}{rcccc}
\toprule
 & \textbf{Depth} & \textbf{States} & \textbf{Memory} & \textbf{Time} \\
 \cmidrule{2-5}
Safety & 23,619,898 & 5.84e+09 & 314 GB & 3.2 hours \\
 \cmidrule{2-5}
consu\_starv & 49,295,820 & 1.01e+10 & 707 GB & 10.4 hours \\
 \cmidrule{2-5}
produ\_starv & 49,678,300 & 1.01e+10 & 707 GB & 10.4 hours \\
 \cmidrule{2-5}
Deadlock & 40,484,755 & 8.11e+09 & 477.2 GB & 6.75 hours \\
\bottomrule
\end{tabular}
\end{table}

Moreover, Spin also provides us an unreached statement list to analyze. The followings reasons are certain for the unreached statements. Note that both priority runqueues are not empty at the scheduling points (never switch to idle thread) in this paper.

\begin{itemize}
\item The scheduling point in PendSV needs to insert current task to the expired runqueue, while the scheduling point in SVC don't. This causes the unreached statements.
\item If the active runqueue always has some elements, the scheduler does not need to perform swap statements and the condition of next thread equals idle will not be checked. This is because there is an enqueue point just before the scheduling point.
\item If the active runqueue is empty and the expired runqueue has some elements, the condition of next thread equals idle will not be checked.
\item If the next task is equal to the current task, need not to perform context switch. This happens only in the pthread yield system call.
\item Two scheduling points in mutex unlock system call never been executed.
\end{itemize}

The last reason is complicated that we need to consider two conditions about the mutex unlock system call. First is the system call from the condition of LL/SC. The second is called via the condition wait system call. In the first situation, consider that the current process has the lock, the thread state must not be \textit{BLOCKED}. The first scheduling point in mutex unlock system call is guarded by the thread state equals \textit{BLOCKED} condition will never be executed under first situation. However, the first scheduling point will always be approached in the second situation. Because the conditional statement in mutex unlock system call is deterministic while the second scheduling point in mutex unlock system call will never be executed. Those are the reason why there are two scheduling point not be executed.

\section{Conclusions}
We adapt an existing machine-assisted framework to fit the requirement of model checking and verify several properties of a simple concurrent program under the specific conditions. The model contains the simulation of hardware and software aspects. Moreover, we do not construct any mathematical proof, the PROMELA language is C-like that anyone can rebuild at any stages of the system life cycle by their own (the effort of this paper is about 7 man-months and the Piko/RT is still in development) and the verification time is within one day (3 hours for safety property). This means model checking have its advantage on verifying a real-world system at the early stage comparing to the system optimized for formal verification.

However, there are still some defects in this paper that need to be accomplished.

\begin{enumerate}
\item The cost (time and memory usage) of adding one process to verify is too expensive in checking the OS source code. There are two possible ways to deal with. First, split the model into several layers and verify each layer one by one. Another way is to construct proof by hand, Delzanno et al. \cite{EPTCS161.13} has proved that using a limited number of process can represent the situation of n processes in Spin.
\item How to claim the abstract model and the real-world system are truly homogeneous. One possible way is to construct a mechanism that extract the PROMELA model from the C source code automatically rather than builds by human.
\end{enumerate}

For the second point, we can not claim the model in this paper is totally correct, yet. But we can tell the behave is similar to the target system by adding \textit{assert}, analyzing the unreached statements or error message, and applying the test cases. In this paper, we use condition variable to deal with the mutual exclusion problem. We also adapted a mutex test case before, and currently, we are trying to deal with the idle thread problem to look inside the behavior of the model.

\bibliographystyle{ACM-Reference-Format}
\bibliography{bibliography}

\end{document}